\documentclass[twocolumn,showpacs,amsmath,amstex,amssymb,mathfonts,prl]{revtex4-1}
\usepackage{amsthm,amsfonts,graphicx,verbatim,times}
\usepackage{amsmath}
\usepackage{amssymb}
\usepackage{amsthm}
\usepackage{amsfonts}
\usepackage{listings}
\usepackage{enumerate}
\usepackage{latexsym}
\usepackage{psfrag}
\usepackage{bm}
\usepackage{graphicx}
\usepackage{subfigure}
\newcommand{\be}{\begin{equation}}
\newcommand{\ee}{\end{equation}}
\newcommand{\bea}{\begin{eqnarray}}
\newcommand{\eea}{\end{eqnarray}}

\newcommand{\ra}{\rangle}

\renewcommand{\phi}{\varphi}
\renewcommand{\epsilon}{\varepsilon}

\begin{document}

\title{Topological Phases in the Zeroth Landau Level of Bilayer Graphene}

\author{Z. Papi\'c$^1$ and D. A. Abanin$^{2,3}$}
\affiliation{$^1$ Department of Electrical Engineering, Princeton University, Princeton, NJ 08544, USA}
\affiliation{$^2$ Perimeter Institute for Theoretical Physics, Waterloo, ON N2L 2Y5, Canada} 
\affiliation{$^3$ Institute for Quantum Computing, Waterloo, ON N2L 3G1, Canada}


\date{\today}

\begin{abstract}
We analyze the phase diagram of the zeroth Landau level
of bilayer graphene, taking into account the realistic effects of screening of the Coulomb
interaction and the strong 
mixing between two degenerate sublevels. We identify robust quantum Hall states
at filling factors $\nu=-1, -\frac{4}{3}, -\frac{5}{3}, -\frac{8}{5}, -\frac{1}{2}$, and discuss the nature of their
ground states, collective excitations, and relation
to the more familiar states in GaAs using a tractable model. In particular,
we present evidence that the $\nu=-\frac{1}{2}$ state, which was recently reported experimentally, is non-Abelian, and described by either the Moore-Read Pfaffian wave function or its particle-hole conjugate, while ruling out other candidates such as the 331 state.
\end{abstract}
\maketitle

{\sl Introduction.} Following a rapid progress in graphene sample quality, the fractional quantum Hall effect (FQHE) was discovered in this material~\cite{Du09,Bolotin09,Dean11,Feldman12,Smet12,Feldman13}. A novel feature of graphene~\cite{CastroNeto09} compared to GaAs-based two-dimensional electron gas (2DEG) is the four-fold degeneracy of Landau levels (LLs) due to spin and valley degrees of freedom. The interplay of long-range SU(4)-symmetric Coulomb interactions and smaller symmetry-breaking terms gives rise to an unusual sequence of FQHEs in the zeroth LL, in which certain states are absent or weak, while others exhibit phase transitions as a function of the magnetic field~\cite{Feldman12,Feldman13,Abanin13}. It was suggested~\cite{Goerbig07,Toke07,papic_ssc, Papic10} that SU(4)-symmetry may give rise to new states not found in GaAs 2DEG and other semiconducting systems (see Ref.~\cite{goerbig_rmp} for a review). 

Very recently, an observation of robust FQHE in the zeroth LL of bilayer graphene (BG) was also reported~\cite{Ki13}. Remarkably, both odd-denominator (filling factor $\nu=-\frac{4}{3}$) and even-denominator ($\nu=-\frac{1}{2}$) fractions were observed. In contrast, monolayer graphene (MG) has so far exhibited only odd-denominator FQHE. What is the nature of the quantum Hall states expected in BG? In particular, is the observed half-integer state an Abelian or a non-Abelian state?

In this Letter, motivated by the experiment~\cite{Ki13}, we provide an insight into the nature of various topological phases arising in the zeroth Landau level of BG. In contrast to previous work~\cite{Shibata09,Apalkov10,Papic11-1,Papic11-2,Apalkov11}, our study fully takes into account the realistic effects of screening of the Coulomb interaction in BG and the strong LL mixing effects between the degenerate sublevels. Using exact diagonalization, we identify robust quantum Hall states corresponding to experimental filling factors $\nu=-1, -\frac{4}{3}, -\frac{5}{3}, -\frac{8}{5}, -\frac{1}{2}$. Furthermore, we introduce a simple model obtained by truncating the Coulomb interaction, which we find to be a helpful guide in understanding the nature of the incompressible ground states and their collective excitations. In particular, we show that odd-denominator states are related to the Laughlin state~\cite{laughlin} and the unprojected composite fermion states~\cite{jain_prb,jainbook}. Furthermore, we present evidence that $\nu=-\frac{1}{2}$ is a non-Abelian state, described by the Moore-Read Pfaffian wavefunction~\cite{Moore91} or its particle-hole conjugate~\cite{apf}, and rule out the 331 state~\cite{halperin} from possible candidates.

{\sl Model.} The effective low-energy Hamiltonian of BG in a magnetic field near $K$-point is given by~\cite{McCann06}: 
\be\label{eq:hamiltonian}
H_{K}=\frac{1}{2m}\left[\begin{array}{cc}
        0 & (\tilde{p}_x+i\tilde{p}_y)^2\\
       (\tilde{p}_x-i\tilde{p}_y)^2 & 0
     \end{array}
 \right] ,\,\, \tilde{\mathbf{p}}=\mathbf{p}+e\mathbf{A},
\ee
where the effective mass is related to the inter-layer tunneling amplitude $\gamma_1$ and the velocity of Dirac excitations in MG, $v_0$, via $m=\gamma_1/2v_0^2$. (The effective Hamiltonian describing excitations near $K'$ point is related to the one above by $H_{K'}=H_K^*$.) The LL spectrum of the Hamiltonian (\ref{eq:hamiltonian}) in the valley $K$ contains two zero-energy LLs with wave functions $(0, |0,m\ra)$, $(0,|1,m\ra)$, where $|n,m\ra$ denotes the wave function in the $n$th non-relativistic LL with angular momentum $m$. In addition, there are non-zero LLs with energies $\epsilon_{n\geq 2}=\pm \hbar \omega_c \sqrt{n(n-1)}$, which are approximately equally spaced at $n\gg 1$. 

We will mostly focus on the vicinity of the filling factor $\nu=0$. In the integer $\nu=0$ state, two pairs of $0,1$ orbitals with the same valley and spin index are filled~\cite{Barlas08,Abanin09}. In this experimentally relevant regime,  spin and valley SU(4) symmetry of the long-range Coulomb interactions is effectively broken down by the short-range valley-anisotropic terms and Zeeman interactions (by an argument similar to the case of MG, see Ref.~\cite{Abanin13}). Recent experimental work~\cite{Maher13} implies that the symmetry breaking is quite strong. Thus, the fractional states in the interval $|\nu|<2$ are likely spin- and valley-polarized. Then, it is sufficient to restrict the Hamiltonian to the 0,1 LL orbitals with a fixed spin and valley index. We note that at $|\nu|>2$ the SU(2) valley symmetry is likely preserved, similar to the case of MG~\cite{Abanin13}. The effects of SU(2) valley symmetry are beyond the scope of this paper (we note, however, that, by analogy with MG, we expect that valley-polarized fractional states will exhibit low-energy skyrmion excitations and will be more susceptible to disorder than their counterparts at $|\nu|<2$). 

\begin{figure}[t]
\includegraphics[width=3.4in]{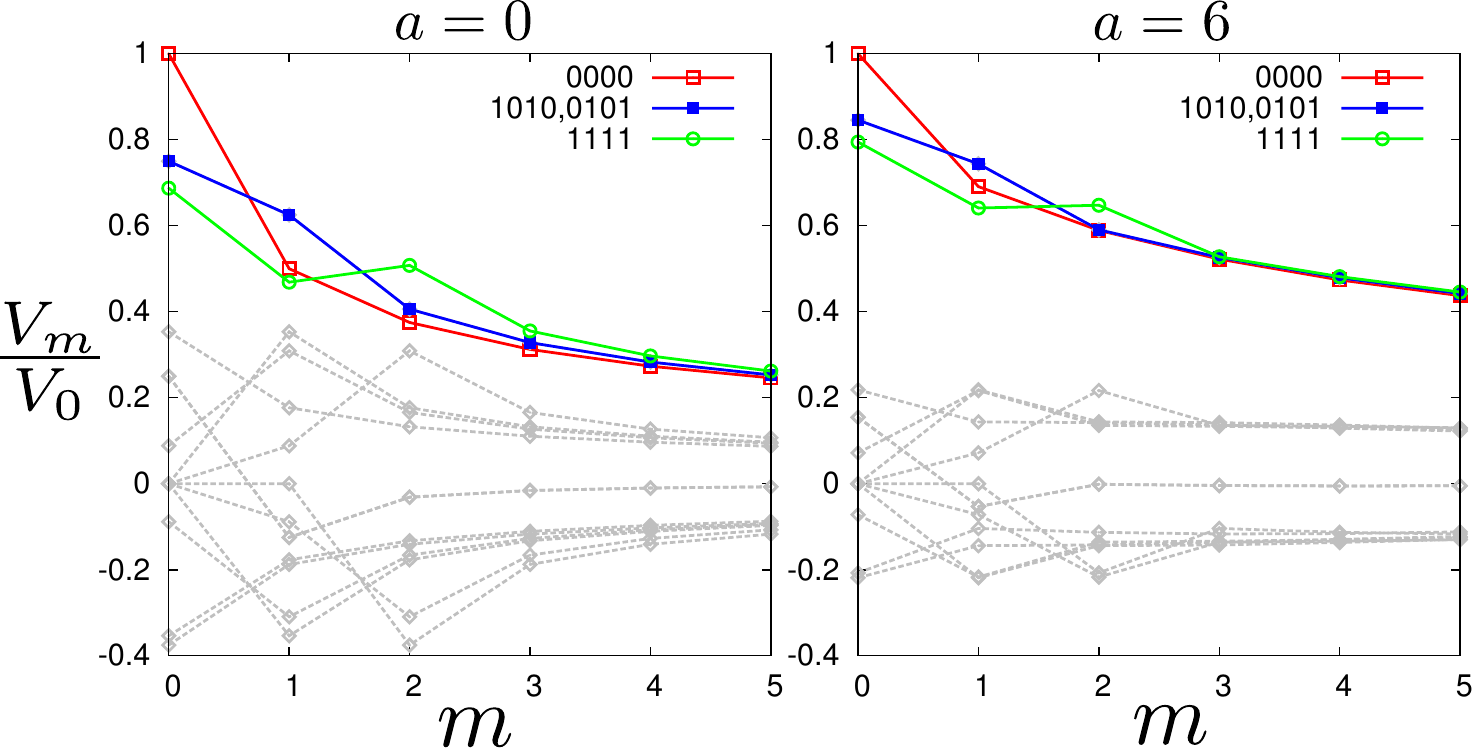}
\vspace{-1mm}
\caption[]{(Color online) Generalized Haldane pseudopotentials for the zeroth LL of BG, for unscreened Coulomb interaction (left), and for screening $a=6$ (right). Values of the pseudopotentials are quoted relative to $V_0^{0000}$. Special combinations of pseudopotentials shown in red, blue, and green define the truncated model (\ref{trunc}), which becomes accurate for large screening. The remaining 12 types of off-diagonal pseudopotentials are shown in gray.}
\label{fig:pseudo}
\vspace{-0pt}
\end{figure}
{\sl Interactions.} The key challenge for the theoretical analysis of FQHE in the zeroth LL of BG lies in the mentioned ``orbital" degeneracy of the zeroth LL due to BG's chiral band structure~\cite{McCann06,Novoselov06}. Since the two LLs (referred to as $0,1$ below)  are exactly degenerate, there is strong mixing between them due to the Coulomb interactions, and considerations based on perturbation theory~\cite{llmix} are not expected to hold in this case.
Thus, the effects of LL mixing have to be accounted for directly, which makes the numerical approaches far more difficult. Because of this challenge, all of previous work~\cite{Shibata09,Apalkov10,Papic11-1,Papic11-2,Apalkov11} was limited to FQHE in the non-zero LLs of BG, which are free of the mentioned orbital degeneracy. 

Another feature of BG that distinguishes it from both MG and the non-relativistic 2DEG, is the significant screening of the Coulomb interactions due to virtual excitations of electron-hole pairs between filled and empty $n\neq 0$ LLs~\cite{Gorbar10,Snizhko12}. In momentum space, the screening function calculated in the random-phase approximation is well described by the following interpolation formula~\cite{Gorbar10}:
\be\label{eq:screening}
V(k)=V_0(k)\frac{k \ell_B}{k\ell_B +a\tanh(b k^2 \ell_B^2)}, \,\, 
\ee
where $\ell_B=\sqrt{\hbar /eB}$ is the magnetic length, $V_0(k)=\frac{2\pi e^2}{\kappa k}$ is the bare Coulomb interaction, and $\kappa$ is the screening constant due to the substrate. Because screening is the strongest for small $k$'s, without loss of generality we fix $b=1/2$. The screening strength is thus defined by the dimensionless parameter $a={4\ln 4} \frac{E_c}{\hbar \omega_c}$, where $E_c=\frac{e^2}{\kappa \ell}$ is the Coulomb energy, and $\hbar\omega_c=\frac{\hbar^2}{m \ell^2}$ is the cyclotron energy. Owing to the smallness of the cyclotron energy in BG ($\hbar \omega_c=2.16[{\rm meV}] B[{\rm T}]$), parameter $a$ is large for experimentally relevant magnetic fields $B\sim 10 {\rm T}$. Thus, screening strongly alters the effective Coulomb interactions. Note that there are also higher order (3-body, etc.) contributions to the screening, which we neglect here.

\begin{figure}[t]
\includegraphics[width=3in]{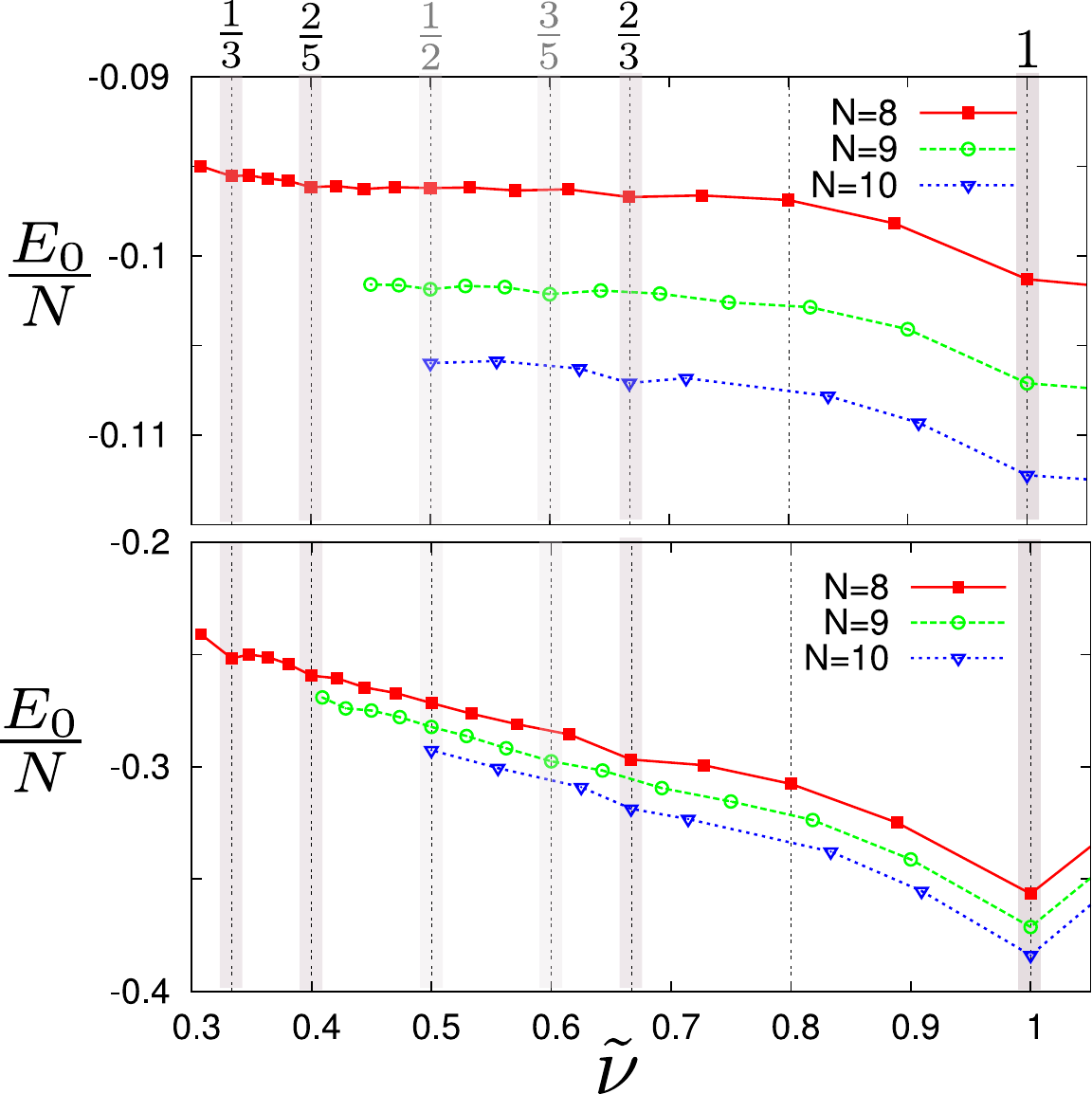}
\vspace{-1mm}
\caption[]{(Color online) Ground state energy per particle as a function of the filling factor in the range $0<\tilde{\nu}<1$, for the full Coulomb interaction (upper panel) and truncated model (lower panel). Shaded regions denote the most robust filling factors $\tilde{\nu}=1, \frac{2}{3}, \frac{2}{5}, \frac{1}{3}$, while weaker features are also observed at $\tilde{\nu}=\frac{1}{2}, \frac{3}{5}$. Data shown is for systems of $N=8,9,10$ particles and screening strength is $a=6$.}
\label{fig:gsenergy}
\vspace{-0pt}
\end{figure}
To quantify the effect of screening, it is useful to project the Coulomb interaction into the space of 0,1 LL orbitals with a given spin/valley polarization. Within this subspace, due to the extra degree of freedom, the Haldane pseudopotentials~\cite{Haldane86,prange} $V_m$ for a pair of particles with a relative angular momentum $m$ must be generalized into
\be\label{eq:pseudo}
V_m^{n_1,...,n_4}=\int\frac{d^2 k}{(2\pi)^2} e^{-k^2} V(k) F_{n1}^{n3}(k) F_{n_2}^{n_4}(-k) F_{m'}^{m}(\sqrt{2}k).
\ee
Here $n_i$ take values 0 and 1, and denote indices of LL orbitals involved in the scattering process defined by the momentum conservation $m'=m+(n_3+n_4)-(n_1+n_2)$. The form-factors are expressed in terms of the generalized Laguerre polynomials $L_{n'}^n$ (see Ref.~\cite{MacDonald94}). 

The first few pseudopotentials for all possible combinations of $\{n_i\}$ are shown in Fig.~\ref{fig:pseudo}. Note that there are linear relations~\cite{unpublished} between different $V_m^{n_1,...,n_4}$, reducing the number of distinct types of $V_m$'s from 16 down to 10. For unscreened Coulomb interaction (Fig.~\ref{fig:pseudo},left), different types of pseudopotentials are of comparable magnitude, making it difficult to understand the role of LL mixing. A more transparent picture emerges in the case of screened interaction (Fig.~\ref{fig:pseudo}, right). In this limit, the diagonal pseudopotentials (i.e., those with $n_1=n_3$, $n_2=n_4$) become dominant, while all the off-diagonal ones become strongly suppressed. This can be easily understood from the real-space structure of the potential (\ref{eq:screening}), $V(r)\propto   \ln (a/r)$, for $1 \ll r/\ell_B \ll a$. The screened potential is thus quite flat in the important interval of distances $1 \ll r/\ell_B \ll a$. Moreover, it is easy to verify that for constant $V(r)=const$, the off-diagonal pseudopotentials identically vanish. Thus, it is natural that flattening of the potential reduces the importance of mixing, as Fig.~\ref{fig:pseudo} illustrates. 

Motivated by these considerations, we introduce the following decomposition of the full Coulomb Hamiltonian 
\begin{equation}\label{trunc}
H_{\rm full} = H_{0000}+H_{1111}+H_{1010}+H_{0101}+ \lambda H_{\rm rest},
\end{equation}
where $\lambda\in [0;1]$. Point $\lambda=0$ defines what we refer to as the ``truncated model", which is useful for developing an intuitive understanding of the ground state in the large-screening limit. Had 0 and 1 LL orbitals had the same form, the truncated model would possess full SU(2) symmetry. Because of the difference in form factors between 0 and 1 LL, this symmetry is slightly broken. In the case of gapped phases, we generally find~\cite{unpublished} that the ground states adiabatically evolve as $\lambda$ is reduced from unity down to zero. We will thus use the truncated model to understand the nature of the ground state when the quantum fluctuations arising from the stronger mixing terms present in $H_{\rm rest}$ are frozen.  

{\sl Results.} In the following, we use exact diagonalization to study the full interacting problem defined by Eqs.(\ref{eq:screening}),(\ref{trunc}), including all types of mixing terms between 0 and 1 LLs. We consider $N$ electrons in an area enclosing $N_\Phi$ flux quanta, assuming fully periodic boundary conditions in isotropic unit cells of high symmetry~\cite{yhl, duncan_translations}. In this case, the filling factor is simply defined as $\tilde{\nu}=N/N_\Phi$. Our convention for $\tilde{\nu}$ below is such that the filling factor of states observed in BG~\cite{Ki13} corresponds to $\nu \equiv -2+\tilde{\nu}$. Generally, we note that in BG, owing to orbital degeneracy and asymmetry between 0 and 1, electron-hole symmetry is broken, and we expect it to be substituted by $\nu\to \nu+2$ symmetry. In addition to the torus, we will also consider spherical geometry~\cite{Haldane86}, which is more convenient for computing overlaps with model wave functions. 

\begin{figure}[t]
\includegraphics[width=3.4in]{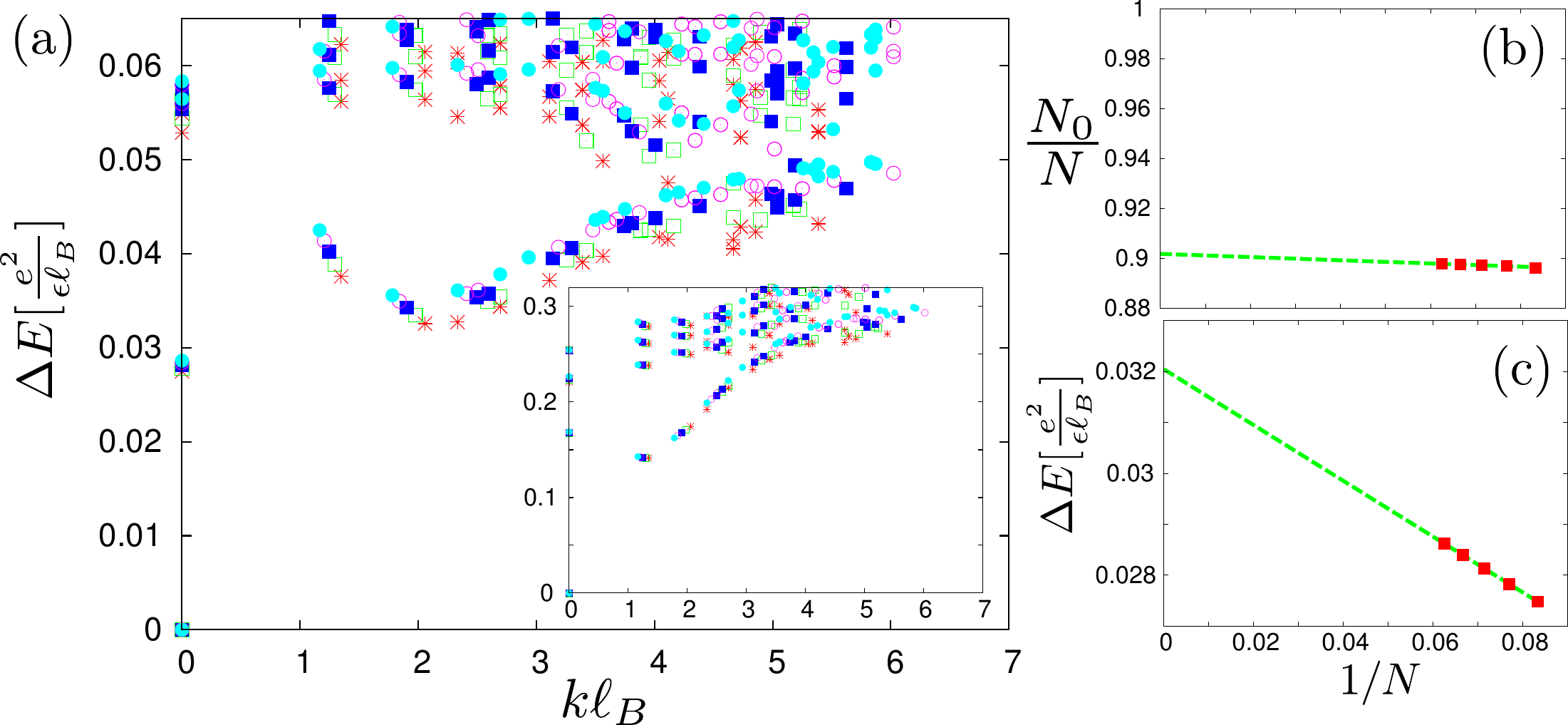}
\vspace{-1mm}
\caption[]{(Color online) Gapped neutral excitations of the $\tilde{\nu}=1$ state. Energy spectrum as a function of momentum for the screened Coulomb interaction $a=6$ (a), and the truncated model (inset). Finite-size scaling of the average population of $n=0$ LL (b) suggests the ground state has a finite component ($\sim 10\%$) in $n=1$ LL. Scaling of the gap is shown in (c). Data is for system sizes $N=12-16$ particles.}
\label{fig:nu1}
\vspace{-0pt}
\end{figure}
Similar to the pioneering work of Yoshioka, Halperin, and Lee~\cite{yhl}, in Fig.~\ref{fig:gsenergy} we first show the plot of the ground state energy per particle, $E_0/N$, as a function of $\tilde{\nu}$. Cusps in $E_0/N$ indicate the presence of gaps for creating charged excitations, and therefore signal the incompressibility. Our analysis shows that this happens at $\tilde{\nu}=1, \frac{2}{3}, \frac{2}{5}, \frac{1}{3}$, which are expected to develop into robust fractions in experiment. As suggested previously, the truncated model (Fig.~\ref{fig:gsenergy},lower) provides a cleaner resolution of the cusps that are more difficult to discern for the full Coulomb interaction (Fig.~\ref{fig:gsenergy},upper), due to finite size effects. Some weak features are also observed at $\tilde{\nu}=\frac{1}{2}, \frac{3}{5}$, but they are size-dependent. Note that curves for different $N$ in Fig.~\ref{fig:gsenergy} collapse onto each other when electrostatic (Madelung) correction is included; we omit such terms in order to enhance the clarity of the cusps.

The strongest feature in Fig.~\ref{fig:gsenergy} is clearly $\tilde{\nu}=1$. Previous work~\cite{Barlas08} has suggested that this state might have unique features that distinguish it from the spinful and bilayer 2DEGs, such as the gapless neutral mode dispersing as $k^{3/2}$. Quite surprisingly, our calculations show that this state is fully gapped in both charge and neutral sectors, Fig.~\ref{fig:nu1}(a),(c). The reason for such discrepancy might be due to the fact that the mean-field ansatz~\cite{Barlas08} assumes a complete polarization of the ground state in $n=0$ LL. As shown in Fig.~\ref{fig:nu1}(b), this is not quite true: in the thermodynamic limit, roughly one out of 10 particles is promoted from $n=0$ into $n=1$ LL. Our truncated model in Eq.(\ref{trunc}) is particularly useful in understanding the physics of the $\tilde{\nu}=1$ ground state and its low-lying excitation spectrum, see Fig.~\ref{fig:nu1}, inset. In the model defined by Eq.(\ref{trunc}), the ground state is indeed a single Slater determinant corresponding to the fully filled $n=0$ LL. The branch of low-lying excitations is also given by single Slater determinants formed by promoting a single particle from $n=0$ into $n=1$, and boosting it with a given momentum $k$. When quantum fluctuations are turned on, these states become dressed with further excitations into $n=1$ LL, e.g. the ground state begins to acquire configurations with 2 particles in $n=1$ LL, the first excitation in $k=0$ sector will contain also 3 particles in $n=1$ LL, etc. The weight of such  configurations can be computed in perturbation theory and will be presented elsewhere~\cite{unpublished}.

\begin{figure}[t]
\includegraphics[width=3in]{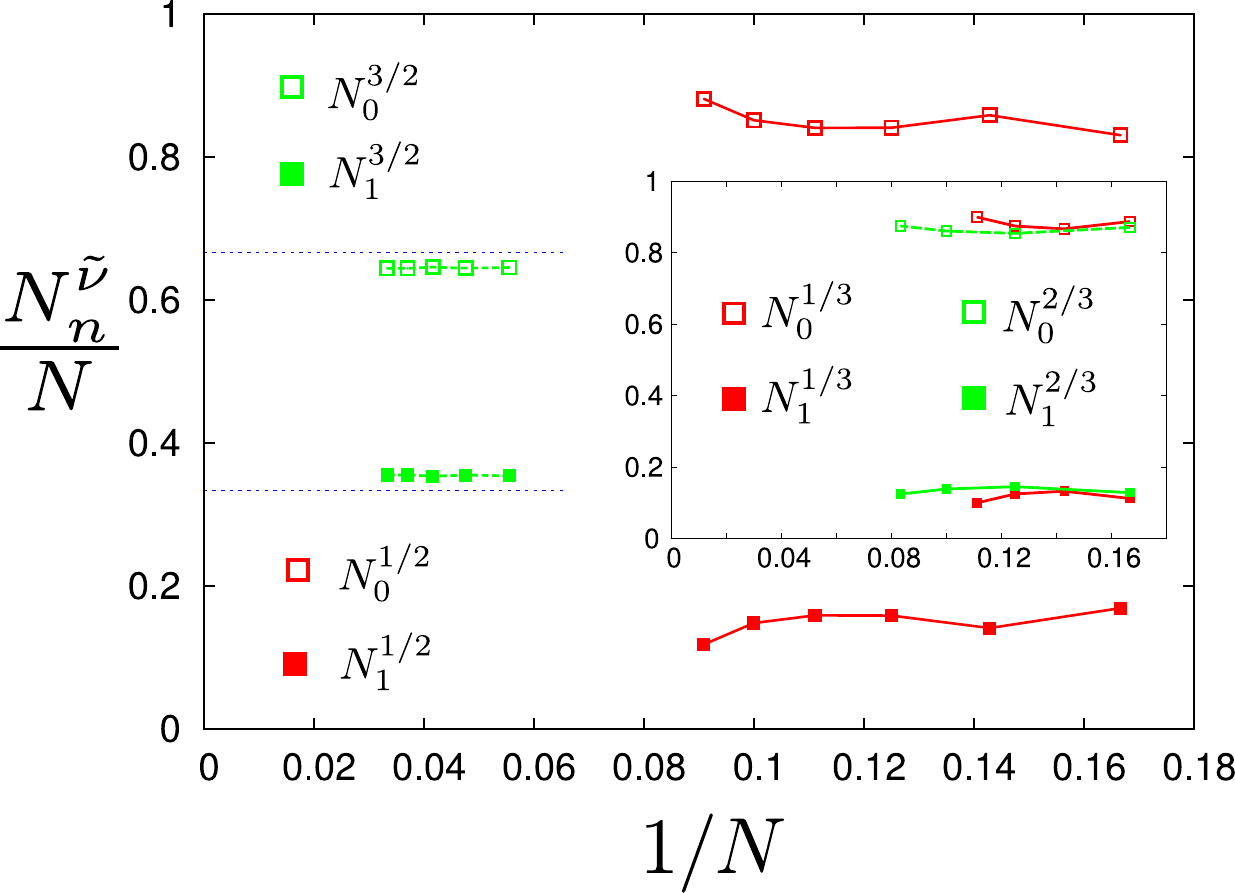}
\vspace{-1mm}
\caption[]{(Color online) Scaling of normalized average populations $N_n^{\tilde{\nu}}$ of the levels $n=0,1$ as a function of system size, for filling factors $\tilde{\nu}=\frac{1}{2},\frac{3}{2}$, and $\tilde{\nu}=\frac{1}{3}, \frac{2}{3}$ (inset). Data is consistent with $\tilde{\nu}=\frac{1}{2},\frac{1}{3},\frac{2}{3}$ being fully polarized in $n=0$ LL. At $\tilde{\nu}=\frac{3}{2}$, the populations converge to $n=0$ LL being completely filled and $n=1$ LL being half filled (dashed lines).}
\label{fig:occupations}
\vspace{-0pt}
\end{figure}
In the remainder, we focus on the fractional states, such as $\tilde{\nu}=\frac{2}{5}, \frac{1}{3}, \frac{2}{3}$, and discuss in particular the case of half-filling, $\tilde{\nu}=\frac{1}{2}, \frac{3}{2}$. At $\tilde{\nu}=\frac{2}{5}$, a compelling candidate wave function is the unprojected Jain state~\cite{jain_prb}, which is defined as a unique and densest zero mode of the $V_1$ Haldane pseudopotential restricted to $n=0,1$ LLs. The overlap between this wave function on the torus and the exact Coulomb ground state is 0.85 ($N=6$) and 0.76 ($N=8$) for screening strength $a=6$, which suggests that the unprojected Jain 2/5 state captures the correct phase, albeit with much stronger corrections than in the well-known cases of \emph{projected} Jain states~\cite{jainbook}.

To understand the nature of half- and third-filled states, it is instructive to first determine the average population of each of the levels, $N_{n=0,1} \equiv \langle \sum_{m} c_{m,n}^\dagger c_{m,n}\rangle$, in the many-body ground state. In Fig.~\ref{fig:occupations} we plot $N_n$ (normalized by $N$) as a function of system size for $\tilde{\nu}=\frac{1}{2},\frac{3}{2}$, as well as $\tilde{\nu}=\frac{1}{3}, \frac{2}{3}$ (inset). While the available system sizes are unfortunately too small to draw definite conclusions, the main trend at $\tilde{\nu}=\frac{1}{2}, \frac{1}{3}, \frac{2}{3}$ fillings suggests that nearly all particles reside in $n=0$ LL in the thermodynamic limit. This is further corroborated in the truncated model, where the ground state is fully polarized in $n=0$ LL in all of these cases and for any finite system. For $\tilde{\nu}=\frac{1}{3}$ and $\tilde{\nu}=\frac{2}{3}$ the gap does not close upon interpolating between the truncated model and the full Coulomb interaction. This suggests that $\tilde{\nu}=\frac{1}{3}$ and $\tilde{\nu}=\frac{2}{3}$ are in the universality class of the usual Laughlin state~\cite{laughlin} and its particle-hole conjugate, with small corrections generated by promoting particles into $n=1$ LL. However, although the gap does not close between the truncated and the full model, we observe several crossings in the excitation spectrum, suggesting that mixing has a stronger effect on the excited states. Indeed, the truncated model predicts the lowest excited states at $\tilde{\nu}=\frac{1}{3},\frac{2}{3}$ to also be fully polarized in $n=0$ LL, hence the two fractions should have identical gaps. In contrast, we find that the gaps for the full Coulomb interaction appear to be $2-3$ larger at $\tilde{\nu}=\frac{2}{3}$ compared to $\tilde{\nu}=\frac{1}{3}$, illustrating that full treatment of the mixing terms appears necessary for quantitative estimates of the gaps.

\begin{figure}[t]
\includegraphics[width=3.4in]{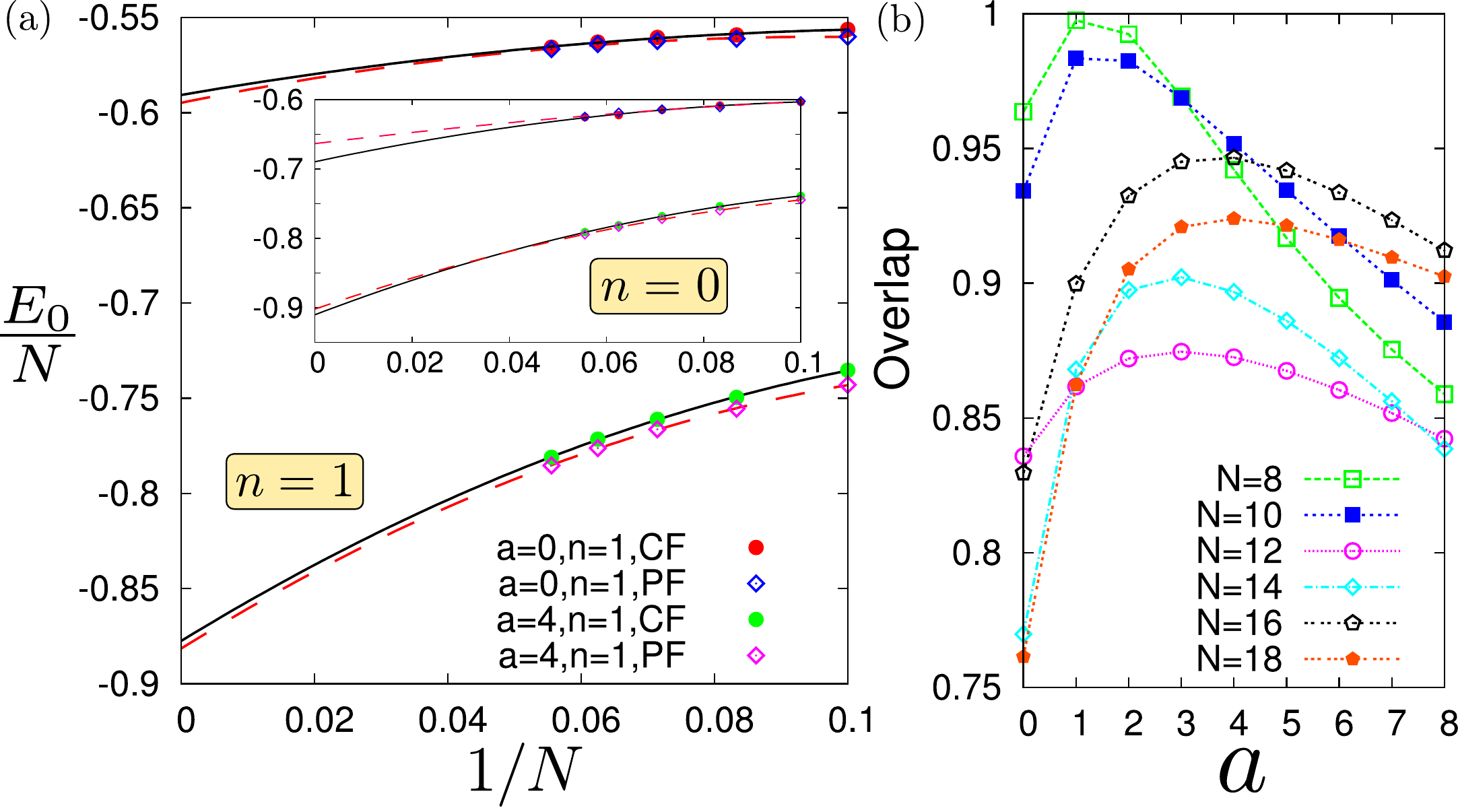}
\vspace{-1mm}
\caption[]{(Color online) Screening favors the paired state over the Fermi liquid. (a) Extrapolated ground-state energies per particle for the shifts corresponding to the Fermi liquid and Pfaffian states on the sphere, in $n=1$ LL and $n=0$ LL (inset). In $n=1$ LL, Pfaffian has a lower energy, making it a better candidate for $\tilde{\nu}=\frac{3}{2}$, while the Fermi liquid has a lower energy at $\tilde{\nu}=\frac{1}{2}$. (b) Screening improves the overlap of $n=1$ ground state with the Pfaffian wave function.}
\label{fig:pfaffian}
\vspace{-0pt}
\end{figure}

On the other hand, full $n=0$ LL polarization observed in the truncated model suggests that $\tilde{\nu}=\frac{1}{2}$ state might be a compressible Fermi liquid state~\cite{HLR}, provided that the effect of screening is not strong enough to cause a phase transition into the incompressible phase. This is consistent with the weak features at $\tilde{\nu}=\frac{1}{2}$ in Fig.~\ref{fig:gsenergy}. Similarly, the populations of levels at $\tilde{\nu}=\frac{3}{2}$ (Fig.~\ref{fig:occupations}) suggest that in this case, $n=0$ LL is likely completely filled, and one might tentatively identify the half-filled $n=1$ LL with the Moore-Read Pfaffian state~\cite{Moore91}. Assuming that polarizations are indeed complete, in Fig.~\ref{fig:pfaffian} we address the effect of screening on the competition between the Fermi liquid and the Pfaffian state in a single isolated ($n=0$ or $n=1$) LL. To this end, we use spherical geometry~\cite{Haldane86}, where the Pfaffian and Fermi liquid states are distinguished by a special quantum number called the shift~\cite{shift}. In Fig.~\ref{fig:pfaffian}(a) we show the scaling of the ground state energy per particle    
with system size, for Pfaffian/Fermi liquid shifts in both $n=0$ and $n=1$ LL. In the latter case, Pfaffian has a lower energy for any value of the screening. Moreover, the overlaps between the exact ground state and the Pfaffian wavefunction are also enhanced by the screening, Fig.~\ref{fig:pfaffian}(b). Combined with the polarization data in Fig.~\ref{fig:occupations}, this evidence points to the non-Abelian bulk physics at $\tilde{\nu}=\frac{3}{2}$.

Due to the peculiarity of half-filling, in addition to the Pfaffian state, its particle-hole conjugate -- the ``anti-Pfaffian"~\cite{apf} -- must also be considered as a possible candidate. These two states are degenerate in the absence of mixing terms, and yield the same overlap with the Coulomb ground state shown in Fig.~\ref{fig:pfaffian}(b). Therefore, to reliably determine which of two is favored, one has to include all mixing terms, as well as higher order (3-body) corrections to the screening. This severely limits the system sizes accessible in the numerics, and lies outside the scope of present work. Note that both Pfaffian and anti-Pfaffian have the same non-Abelian physics in the bulk, and differ mainly in their edge excitations~\cite{apf}.  

From Fig.~\ref{fig:pfaffian}(b) it is obvious that screening has a favorable effect on the pairing correlations at half-filling. In the case of $\tilde{\nu}=\frac{1}{2}$, the Fermi liquid energy clearly lies lower than that of the Pfaffian for small/zero screening, but they approach each other for larger values of the screening. It thus remains possible that large screening leads to a destruction of the Fermi liquid phase. However, we believe that the resulting phase would nevertheless have a much smaller gap than at $\tilde{\nu}=\frac{3}{2}$.

Finally, it is often speculated that 331 state~\cite{halperin} might represent an alternative candidate for the incompressible phase at half filling. Our calculations have neglected the real electron spin, therefore the 331 state can only occur as a ``pseudospin" state, with $n=0$ and $n=1$ LLs playing the role of $\uparrow$, $\downarrow$ spins. However, the correlations built into the 331 wave function enforce the populations $N_0$ and $N_1$ to be equal, and a small imbalance between $N_0$ and $N_1$ quickly destroys the state~\cite{pgrm,ppds}. Thus, we believe this state is not a viable candidate at $\tilde{\nu}=\frac{3}{2}$ where $N_0$ is nearly twice as large as $N_1$ (Fig.~\ref{fig:occupations}). Note that one could imagine generalizations of the 331 state that appear better adapted to the BG problem. For example, one could take the $V_1-V_3$ model which is the parent Hamiltonian of the usual 331 state, and multiply it with the appropriate form factors of the $0,1$ LLs. At $\tilde{\nu}=\frac{3}{2}$, the ground state of such a Hamiltonian also fully fills $n=0$ LL, but its component in $n=1$ LL has a small overlap with that of the Coulomb ground state, which is still further suppressed by screening. Thus 331-based wave functions appear to be unlikely candidates for $\tilde{\nu}=\frac{3}{2}$ in BG. 

In conclusion, we have presented a method to analyze the effect of screening and strong LL mixing in the zeroth LL of BG. We have identified several robust fractions with Abelian ($\nu=-1, -\frac{4}{3}, -\frac{5}{3}, -\frac{8}{5}$) and non-Abelian ($\nu=-\frac{1}{2}$) topological order, some of which have been observed in recent experiments~\cite{Ki13}. Future work will address the microscopic characterization of the excitations in these states, and ways to further  increase their gaps or perhaps stabilize new states using the potential  tunability of the interactions in BG~\cite{tunable, Papic11-1, Papic11-2}.  

{\sl Acknowledgements}. We thank B. Feldman, B. Halperin, A. Kou, D.-K. Ki, A. Morpurgo, and A. Yacoby for useful discussions. This work was supported by DOE grant DE-SC$0002140$.

\end{document}